\documentclass[12pt]{article}

\usepackage{scicite}

\usepackage{times}

\usepackage{amsmath}

\usepackage{amssymb}

\usepackage{graphicx}

\usepackage{url}

\usepackage{comment}

\usepackage{xr}
\usepackage[usenames]{color}

\usepackage{rotating}
\usepackage{color}

\usepackage{upgreek}
\usepackage{array}
\usepackage{subfigure}

\newcolumntype{M}[1]{>{\centering\arraybackslash}m{#1}}

\usepackage{lineno,xcolor}

\newenvironment{sciabstract}{%
\begin{quote} \bf}
{\end{quote}}

\newcounter{lastnote}

\title{Mileage efficiency and relative emission of automotive vehicles}

\author
{Neelesh A. Patankar,$^{1\ast}$ Tanvee N. Patankar$^{2}$\\
\\
\normalsize{$^{1}$Department of Mechanical Engineering, Northwestern University, Evanston, IL 60208, USA}\\
\normalsize{$^{2}$Adlai E. Stevenson High School, Lincolnshire, IL 60069, USA}\\
\\
\normalsize{$^\ast$To whom correspondence should be addressed; E-mail:  n-patankar@northwestern.edu.}
}

\date{}

\begin{document} 


\baselineskip24pt


\maketitle


\begin{sciabstract}

Physics dictates that cars with small mass will travel more miles per gallon (mpg) compared to massive trucks. Does this imply that small cars are more efficient machines? In this work a mileage efficiency metric is defined as a ratio of actual car mileage (mpg) to the mileage of an ideal car. This metric allows comparison of efficiencies of cars with different masses and fuel types. It is as useful to quantify efficiencies of cars as the concept of drag coefficient is to quantify the efficacy of their aerodynamic shapes. 
Maximum mileage and lowest $\textup{CO}_2$ emission of conventional gasoline cars, at different driving schedules, is reported based on the concept of an ideal car. 
This can help put government imposed standards in a rigorous context. 
  
\end{sciabstract}




\clearpage

\section*{Introduction}

The world record for maximum mileage is claimed to be 12,665 miles per gasoline gallon equivalent (MPGe) set by PAC--Car II \cite{pacc}. This seems astonishing because there is no known meaningful reference point in the form of an idealized car mileage or an upper limit for mileage with which this record can be compared. It will be shown here that a hypothetical car with an ideal powertrain but with same road load losses (aerodynamic drag and rolling friction) as that of PAC--Car II, would give 75,172 MPGe. This implies PAC-Car II mileage is 16.8$\%$ of this idealized limit. In light of this result, the high mileage of PAC--Car II is not as astonishing. On the contrary, typical mileages ($\sim$ 30 mpg) of conventional (i.e. without regenerative braking) fossil fuel cars now seem more surprising. 

What is the mileage efficiency of the cars we drive? One goal of this work is to define a mileage efficiency metric for automotive vehicles. This metric will be used to compare performances of vehicles of different masses. For example, by laws of physics, a car with small mass will give greater mileage compared to a massive truck. Does this mean that the small car is a more efficient machine? It will be shown that average efficiencies of cars of different masses are not significantly different. 

Mileage efficiency, which is a non-dimensional measure, is formally defined in this work as the ratio of actual car mileage (MPGe) to the mileage of an ideal car. Similarly, a non-dimensional measure of fuel consumption is formally defined as the ratio of fuel consumption of a real-world car to the fuel consumption of an ideal car. This ratio will be called the energy consumption coefficient ($C_E$). Mileage efficiency and energy consumption coefficient metrics allow comparison of efficiencies of different vehicles. Our proposition is that these metrics are as useful to quantify efficiencies of cars as the concept of drag coefficient ($C_d$) is to quantify the efficacy of their aerodynamic shapes.   


Meaningful ways to define an ideal car will be proposed in this work. It will be shown how the mileage efficiency can be computed from the knowledge of efficiencies of various components of cars. 
No claim is made that an ideal car would be practically feasible. However, it allows putting mileages of real-world cars in perspective similar to how efficiency of a machine informs us how well that machine is functioning relative to an ideal limit. Finally, in this work only conventional cars defined as those with no regenerative braking will be considered. Generalization of this approach to hybrid and other types of vehicles has no fundamental conceptual barrier and merits future investigation. The practical utility of the concept of mileage efficiency and energy consumption coefficient is demonstrated. 


\section*{Ideal cars}

Consider a car that is cruising at constant speed on a level road. If one considers ideal conditions which implies 100$\%$ efficiency of the internal components of the car and no external losses in the form of aerodynamic drag and rolling friction, then by Newton's first law of motion this ideal car should require no energy to sustain it's constant speed -- it would give ``infinite" miles per gallon.

Now consider the same ideal car, as above, on a level road but undergoing a driving schedule that includes acceleration and deceleration phases. During acceleration, by Newton's second law, the car would require power input to increase it's kinetic energy. Suppose that braking is used to decelerate it and that the braking energy is fully recovered by way of a 100$\%$ efficient ideal regenerative brake. Additionally, there is no power loss during idling of this ideal car. If there is no net change in kinetic energy in the driving schedule, then this ideal car too would require no net energy to move -- once again implying ``infinite" miles per gallon. 

Aforementioned examples make it clear that an ideal car defined in the above sense, which will be called an \emph{ideal Newton car}, would give ``infinite" miles per gallon on a level road. Comparing real car mileage with that of an ideal Newton car is not particularly useful or insightful. Hence, two other types of ideal cars are defined below that provide a meaningful reference to compare the mileage of real-world cars.
  
The first type of ideal car is defined as the one where there are no losses in the powertrain or idling and there is no aerodynamic drag or rolling friction. The powertrain includes all components from fuel to wheel traction. An ideal powertrain implies that all the power in the fuel is delivered in the form of tractive power. In this type of ideal car, the conditions are ideal in every way except that during deceleration the kinetic energy is lost as heat by way of brakes. Thus, this ideal car is a conventional car with brakes. There is no regenerative braking. This ideal car will be called an \emph{ideal brake-loss car} (IBC). An ideal brake-loss car will require no energy during cruising, it will require power during acceleration or to go up a grade, and it will loose energy to heat during deceleration by way of braking. 

Few comments are in order. For cruising driving schedules where there is no acceleration or deceleration, an ideal brake-loss car will give ``infinite" miles per gallon. However, for a typical driving schedule of real-world cars that are dominated by acceleration and deceleration phases, an ideal brake-loss car will have finite mileage. The mileage of an ideal brake-loss car will depend, as expected, on the driving schedule, among other factors. Finally, static friction is considered strong enough to cause pure rolling of tires at all times so that there is no sliding friction loss. The mileage of an ideal brake-loss car will be the upper limit of mileage for cars that do not have regenerative braking. However, regenerative braking can help achieve mileages greater than an ideal brake-loss car provided there is no other energy loss for the car. 

The second type of ideal car is defined as the one that has an ideal powertrain, which includes no fuel consumption during idling. There is energy loss via braking. Additionally, road load losses in the form of aerodynamics drag and rolling resistance are the same as the real-world car to which this ideal car would be compared. This ideal car will be called an \emph{ideal-powertrain car} (IPC). Note that unlike an ideal brake-loss car, an ideal-powertrain car will have finite mileage for cruising driving schedules.

The choice IBC and IPC as two types of ideal cars is not arbitrary. It will be apparent from mileage efficiency equations, to be derived below, that these two ideal cars represent meaningful references to compare the mileages of conventional real-world cars. 

\section*{Ideal car mileage}

First an expression for ideal brake-loss car mileage will be derived and then an expression for the mileage efficiency of a real-world car will be derived. Level road will be considered in the following analysis, however, inclusion of the grade can be done without any fundamental difficulty.

Consider a driving schedule in which the total distance traveled is $D$ miles. Let $E_{a+}$ be the total energy required during the acceleration phases in the driving schedule. $E_{a+}$ can be computed as
\begin{equation}
E_{a+} = \sum\limits_{i}\left(~{\int\limits_{acc_i}{M \frac{dV}{dt}Vdt}}\right)
= \sum\limits_{i}\left(~{\int\limits_{acc_i}{\frac{d}{dt}\left(\frac{1}{2}MV^2\right)dt}}\right),
\label{eqn-p-acc}
\end{equation}
where $M$ is the mass of the vehicle and $V$ is the speed of the vehicle. The integral is over an $i^{th}$ acceleration phase of the schedule and the summation is over all acceleration phases $i$. The total energy lost during deceleration, $E_{a-}$, is equal to $E_{a+}$ in a driving schedule that has no net gain or loss of kinetic energy. An ideal brake-loss car that has no aerodynamic drag or rolling friction would have to decelerate entirely by way of braking. Thus, for an ideal brake-loss car:
\begin{equation}
E_{b,IBC} = E_{a-} = E_{a+},
\label{eqn-brake-ibc}
\end{equation} 
where $E_{b,IBC}$ is the energy lost by an ideal brake-loss car during braking. Real-world cars would loose some of the deceleration energy to aerodynamic drag and rolling friction \cite{Sovr83a}. Thus, the braking energy of an ideal brake-loss car would be larger than that of a real-world car for the same driving schedule. It follows that $E_{b,IBC}$ is the maximum available energy for regeneration by brakes if regenerative braking is used. For an ideal brake-loss car the energy required to move through a driving schedule is $E_{a+}$. Due to ideal assumption this would be equal to the energy of the fuel used
\begin{equation}
E_{f,IBC} = E_{a+},
\label{eqn-fuel-ibc}
\end{equation} 
where $E_{f,IBC}$ is the fuel energy that is used for the driving schedule.

The mileage in miles per gallon equivalent (MPGe) of an ideal brake-loss car is given by
\begin{equation}
\textup{MPGe}_{IBC} = \frac{D E_{geg}}{E_{a+}},
\label{eqn-mileage-ibc}
\end{equation} 
where $E_{geg}$ is the energy in a gasoline-equivalent gallon of the fuel (i.e. the energy in one gallon of gasoline). According to the United States Environmental Protection Agency (EPA), $E_{geg} = $ 33.7 kWh of energy \cite{egeg12}.  The fuel consumption,  $\textup{fc}_{IBC}$, in gallons equivalent per mile for an ideal brake-loss car is given by
\begin{equation}
\textup{fc}_{IBC} = \frac{1}{\textup{MPGe}_{IBC}} = \frac{E_{a+}}{D E_{geg}}.
\label{eqn-fc-ibc}
\end{equation} 

The mileage of an ideal brake-loss car depends only on the driving schedule and the mass of the car. For urban (Federal Test Procedure or FTP) and highway (Highway Fuel Economy or HWFET or HWY) schedules from EPA \cite{drivesched}, the ideal brake-loss car mileages are found to be
\begin{equation}
\textup{MPGe}_{IBC} = \begin{cases}
\frac{430772}{M}, & \textup{ for the urban schedule,} \\
\frac{1068645}{M}, & \textup{ for the highway schedule,}
\end{cases}
\label{eqn-ibc-sched}
\end{equation}
where $M$ is in kg. This implies that, for an urban schedule, an ideal brake-loss car mileage of a typical mid-size 2000~kg car is 234~MPGe and the mileage of a fully loaded 36,000~kg semi-truck is 13~MPGe. For a highway schedule the mileages are 580~MPGe and 31~MPGe for a 2000~kg car and a 36,000~kg semi-truck, respectively. For a car or a truck \emph{with no regenerative braking technology}, these ideal brake-loss car mileages are the upper limits, for the respective schedules, that are imposed by constraints of physics no matter what type of fuel technology is used. These mileages can be used to define energy consumption coefficient and mileage efficiency of real-world cars. 

\section*{Energy consumption coefficient}

In this section, the goal is to compare the fuel consumption of a real-world car to that of an ideal car. Models for fuel consumption of real-world cars have been developed~\cite{Simp03a, Simp05a, bagl07a, bagl07b, Berr07a, Band08a, Delo09a, chea10a, chea11a, Halb10a, Alle12a, Alle13a}. The equation of motion of a car can be written according to Newton's second law 
\begin{equation}
M\frac{dV}{dt} = F_p - F_{rl}, ~\textup{i.e.}, ~ F_p = M\frac{dV}{dt} + F_{rl},
\label{eqn-of-motion}
\end{equation} 
where $F_p$ is the total propulsion force at the wheels and $F_{rl}$ is the road load that includes aerodynamic drag and rolling friction.
For simplicity a level road is considered and the wheel rotational inertia is ignored. Terms corresponding to grade and wheel rotational inertia can be added without loss of generality. 
Eqn.~\ref{eqn-of-motion} shows that propulsion power ($=F_pV$) must be provided to accelerate the car and to overcome road load. The propulsion power is positive ($V$ is always positive in this analysis) during acceleration. In this case energy must be expended by the car. Even during deceleration, the propulsion power can be positive due to road load -- this is termed powered deceleration \cite{Sovr83a}. When the propulsion power is negative, brakes must be applied. 

The energy, $E_r$, required for a driving schedule is given by
\begin{equation}
E_r = E_{a+} + E_{rl,a+} + E_{rl,a0} + E_{pa-},
\label{eqn-reqd}
\end{equation}
where $E_{rl,a+}$ is the energy required to overcome road load (aerodynamic drag and rolling resistance) during acceleration, $E_{rl,a0}$ is the energy required to overcome road load during cruising, and $E_{pa-}$ is the propulsive energy required during powered deceleration. The actual fuel energy $E_f$ expended during this schedule is given by 
\begin{equation}
E_f = \frac{E_r}{\eta_p} + E_{v0} = \frac{1}{\eta_p}\left[E_{a+} + E_{rl,a+} + E_{rl,a0} + E_{pa-}  \right] + E_{v0},
\label{eqn-energy}
\end{equation}
where $\eta_p$ is an overall powertrain efficiency for the non-idling part of the driving schedule and $E_{v0}$ is the energy consumed due to fuel spent during idling at zero velocity. Additional power consumption such as air-conditioning could also be added in this term.
Dividing by $DE_{geg}$ gives an expression for fuel consumption $\textup{fc}$ in gallons equivalent per mile (gepm)
\begin{equation}
\textup{fc} = \frac{1}{\eta_p}\left[\textup{fc}_{IBC} + \textup{fc}_{rl,a+}+\textup{fc}_{rl,a0}+ \textup{fc}_{pa-} \right] +  \textup{fc}_{v0} ,
\label{eqn-fc-real}
\end{equation}
where $\textup{fc}_{rl,a+}$ and $\textup{fc}_{rl,a0}$ are the fuel consumptions required to overcome road loads during acceleration and cruising, respectively. $\textup{fc}_{pa-}$ is the fuel consumption required during powered deceleration and $\textup{fc}_{v0}$ is the fuel spent during idling. 

An energy consumption coefficient $C_E$ is defined as the fuel consumption normalized by the fuel consumption of an ideal brake-loss car:
\begin{equation}
C_E = \frac{\textup{fc}}{\textup{fc}_{IBC}} = \frac{1}{\eta_p} \left[ 1 + C_{rl,a+} +  C_{rl,a0} + C_{pa-} \right] +  C_{v0} ,
\label{eqn-C-E}
\end{equation}
where $C_{rl,a+} = \frac{\textup{fc}_{rl,a+}}{\textup{fc}_{IBC}}$ and $C_{rl,a0} = \frac{\textup{fc}_{rl,a0}}{\textup{fc}_{IBC}}$ are the energy consumption coefficients due to road load during acceleration and cruising, respectively. $C_{pa-} = \frac{\textup{fc}_{pa-}}{\textup{fc}_{IBC}}$ is the energy consumption coefficient due to powered deceleration and $C_{v0} = \frac{\textup{fc}_{v0}}{\textup{fc}_{IBC}}$ is the idling energy consumption coefficient. 

$C_E$ depends on various energy loss mechanisms of the car. Internal losses are represented by the non-idling powertrain efficiency $\eta_p$ and idling loss $C_{v0}$, whereas external road load losses due to aerodynamic drag and rolling friction are accounted by $C_{rl,a+}$, $C_{rl,a0}$, and $C_{pa-}$. If there are no internal or external losses, then the car is an ideal brake-loss car and $C_E$ is equal to one. The lower the value of $C_E$, the better the car. For a conventional car that has no regenerative braking, $C_E = 1$, corresponding to an ideal brake-loss car, is the lower bound. 

\section*{Mileage efficiency}

Mileage efficiency with respect an ideal brake-loss car is given by   
\begin{equation}
\textup{MPGe}_{\%IBC} =100 ~\frac{\textup{MPGe}}{\textup{MPGe}_{IBC}} = 100~\frac{1}{C_E} = 100~\frac{ \eta_p}{\left[ 1 + C_{rl,a+} +  C_{rl,a0} + C_{pa-} \right] + \eta_p C_{v0}}.
\label{eqn-MPGe-IBC-pc}
\end{equation}
Eqn.~\ref{eqn-MPGe-IBC-pc} shows that the mileage efficiency with respect to an ideal brake-loss car depends on internal losses ($\eta_p$ and $C_{v0}$) and external losses ($C_{rl,a+}$, $C_{rl,a0}$, and $C_{pa-}$). In fact, for conventional cars without regenerative braking $\textup{MPGe}_{\%IBC}$ cannot exceed $100\%$, thus reinforcing the idea that an ideal brake-loss car represents an upper limit for mileage.

Mileage efficiency with respect to an ideal-powertrain car is obtained as follows. The fuel consumption, $\textup{fc}_{IPC}$, of an ideal-powertrain car can be obtained from Eqn.~\ref{eqn-fc-real} with $\eta_p = 1$ and $\textup{fc}_{v0} = 0$: 
\begin{equation}
\textup{fc}_{IPC} = \textup{fc}_{IBC} + \textup{fc}_{rl,a+}+\textup{fc}_{rl,a0}+ \textup{fc}_{pa-}.
\label{eqn-fc-IPC}
\end{equation}
Similarly, the energy consumption coefficient $C_{E,IPC}$ of an ideal-powertrain car can be obtained from Eqn.~\ref{eqn-C-E} with $\eta_p = 1$ and $C_{v0} = 0$:
\begin{equation}
C_{E,IPC} = \frac{\textup{fc}_{IPC}}{\textup{fc}_{IBC}} = 1 + C_{rl,a+} +  C_{rl,a0} + C_{pa-} = \frac{1}{\textup{RLA}},
\label{eqn-C-E-IPC}
\end{equation} 
where RLA as defined above is Road Load Attrition. $\textup{RLA} = 1$ when there is no drag or rolling resistance (ideal case) whereas RLA is less than one for real cases. Eqns.~\ref{eqn-fc-real} and~\ref{eqn-fc-IPC} imply that
\begin{equation}
\textup{MPGe}_{\%IPC} =100 ~\frac{\textup{MPGe}}{\textup{MPGe}_{IPC}} = 100~\frac{\textup{fc}_{IPC}}{\textup{fc}} 
= 100~\frac{\eta_p}{1+\eta_p C_{v0} \textup{RLA}}.
\label{eqn-MPGe-IPC-pc}
\end{equation}
Furthermore, Eqns.~\ref{eqn-MPGe-IBC-pc},~\ref{eqn-C-E-IPC}, and~\ref{eqn-MPGe-IPC-pc} imply that
\begin{equation}
\textup{MPGe}_{\%IBC} = \frac{\textup{MPGe}_{\%IPC}}{\left[ 1 + C_{rl,a+} +  C_{rl,a0} + C_{pa-} \right]}
=\textup{RLA}~\textup{MPGe}_{\%IPC},
\label{eqn-MPGe-IBC-IPC-pc}
\end{equation}
where Road Load Attrition (RLA) reduces the mileage of a car relative to the ideal brake-loss car due to the road load. Thus, $\textup{MPGe}_{\%IPC}$ primarily captures the effect of internal losses on mileage loss whereas $\textup{MPGe}_{\%IBC}$ captures the effect of internal and external losses on mileage loss. This further highlights the utility of defining ideal brake-loss and ideal-powertrain cars as useful idealizations to compare the mileages of real-world cars.

\section*{Ideal and relative $\textup{CO}_2$ emission} 

First consider the emission of an ideal brake-loss car. 
By definition we require that an ideal car would burn hydrocarbon fuels like gasoline and diesel completely to create only carbon dioxide and water in their exhaust. Let $\Gamma_{idl}$ be the amount of $\textup{CO}_2$ emitted from a completely burned fuel. The EPA 
specifies $\Gamma_{idl}=$ 8,887 grams of $\textup{CO}_2$ emission per gallon of gasoline consumed and 10,180 grams of $\textup{CO}_2$ emission per gallon of diesel \cite{gco2}. Since 0.88 gallon diesel is equivalent to 1 gallon of gasoline \cite{gge}, it implies $\Gamma_{idl} =$ 8,958 grams of 
$\textup{CO}_2$ emission per gasoline-equivalent gallon ($\textup{g}\textup{CO}_2/\textup{geg}$) of diesel. We calculated 6,190 grams of $\textup{CO}_2$ emission per gallon of 85\% ethanol, which is consistent with data in literature \cite{gco2ethanol_1, gco2ethanol_2}. Since, 1.39 gallons of 85\% ethanol is equivalent to a gallon of gasoline \cite{gge}, it implies $\Gamma_{idl} =$ 8,604 $\textup{g}\textup{CO}_2/\textup{geg}$. 

The emission $\zeta_{IBC}$ of an ideal brake-loss car is given by
\begin{equation}
\zeta_{IBC} = \textup{fc}_{IBC}~\Gamma_{idl}.
\label{eqn-ideal-emission}
\end{equation}
$\zeta_{IBC}$ is in $\textup{g}\textup{CO}_2$ per mile ($\textup{g}\textup{CO}_2/\textup{mi}$) and represents the lower limit of emission, given that the fuel is completely burned, for the fuel type corresponding to $\Gamma_{idl}$ used in Eqn.~\ref{eqn-ideal-emission}.
  
Now consider real-world cars. The car engine does not burn the fuel perfectly. In the process of burning gasoline or other hydrocarbon fuel imperfectly, in addition to $\textup{CO}_2$, other gases like carbon monoxide (CO), nitrogen oxides (NOx), and unburned hydrocarbons are produced. Catalytic convertors convert CO, NOx, and hydrocarbons to produce $\textup{CO}_2$, nitrogen, water, and oxygen. 
The total greenhouse gas emission can be quantified in terms of carbon dioxide equivalency (CDE). CDE is a quantity that describes, for a given amount of greenhouse gas, the amount of $\textup{CO}_2$ that would have the same global warming potential (GWP), when measured over a specified timescale (generally, 100 years). Thus, all emissions from a real-world car can be quantified by an overall emission $\zeta$ in grams carbon dioxide equivalent per mile ($\textup{g}\textup{CDE}/\textup{mi}$). The emission of real-world cars can be quantified relative to the ideal emission. The relative emission $\gamma_e$ is given by
\begin{equation}
\gamma_e = \frac{\zeta}{\zeta_{IBC}} = \frac{\textup{fc}~\Gamma}{\textup{fc}_{IBC}~\Gamma_{idl}} =C_E \frac{\Gamma}{\Gamma_{idl}} = \frac{C_E}{\eta_r},
\label{eqn-relative-emission}
\end{equation} 
where fc is the fuel consumption of a real-world car (Eqn.~\ref{eqn-fc-real}) and $\Gamma$ in units of $\textup{g}\textup{CDE}/\textup{geg}$ is the actual CDE emission in the exhaust per gasoline-equivalent gallon of fuel burned. $C_E$ follows from the definition in Eqn.~\ref{eqn-C-E}. $\eta_r = \frac{\Gamma_{idl}}{\Gamma}$, defined in Eqn.~\ref{eqn-relative-emission}, is a measure of the combined efficiency of the combustion process and the catalytic convertor; its value will typically range between 0 to 1. The relative emission $\gamma_e$ will have a lower limit of 1 for conventional internal combustion engine cars. It follows from Eqn.~\ref{eqn-relative-emission} that $\gamma_e$ depends on the reaction efficiency $\eta_r$ (combustion and catalytic convertor) and the mechanical efficiency quantified by $C_E$ (internal and external losses).

\section*{Results} 

In this section the utility of the metrics defined above is demonstrated.

\subsection*{Mileage efficiency of PAC-Car II} 

PAC-Car II reported a mileage of 12,665 MPGe in 2005 during the Shell Eco-Marathon competition. While the exact driving schedule of the car is not available, it known that during the competition, cars had to attain an average speed of at least 15 mph over a distance of 10 miles. The course was a motor racing track. Hence, for simplicity a constant speed driving schedule is assumed for the car due to lack of additional information. As noted earlier, the ideal brake-loss car would give infinite miles per gallon (i.e. requires no energy) at constant speed according to Newton's first law of motion. Consequently, in this case, the ideal-powertrain car, where the powertrain is assumed to be ideal but air and rolling resistances are present, will be used for mileage comparison with PAC-Car II. The goal is to find $\textup{MPGe}_{\%IPC}$ (Eqn.~\ref{eqn-MPGe-IPC-pc}).

PAC-Car II had an average speed of 18.6 mph over a 12.9 mile distance. The air and rolling resistances are quantified by the drag coefficient $C_d =$ 0.075 and the rolling resistance coefficient $\mu_r =$ 0.0008, respectively. The car was extremely lightweight (29 kg) with a frontal area $\textup{A}_f = 0.254~\textup{m}^2$. The fuel consumption, $\textup{fc}_{IPC}^{\textup{PAC}}$, of an ideal-powertrain car with the same air and rolling resistance as PAC-Car II is computed as follows
\begin{equation}
\textup{fc}_{IPC}^{\textup{PAC}} = \textup{fc}_{rl,a0} 
= \frac{1}{E_{geg}} \left(\frac{1}{2}\rho V^2 \textup{A}_f~C_d + \mu_r M g \right),
\label{eqn-fc-IPC-Pac}
\end{equation} 
where only $\textup{fc}_{rl,a0}$ in Eqn.~\ref{eqn-fc-IPC} is non-zero in this problem, $\rho$ ($= 1.2~\textup{kg/m}^3$) is the density of air, and $g$ ($= 9.81~\textup{m/s}^2$) is the gravitational acceleration. The mileage of the ideal-powertrain car is the reciprocal of fuel consumption $\textup{fc}_{IPC}^{\textup{PAC}}$ and it is found to be 75,172 MPGe by using the parameters listed above. This implies that a car with the same air and rolling resistance but with an ideal powertrain would go 75,172 miles in a gallon of gasoline. This provides an idealized reference mileage of PAC-Car II.

The mileage efficiency of PAC-Car II relative to an ideal-powertrain car is $\textup{MPGe}_{\%IPC}^{\textup{PAC}} = 100~\frac{12,665}{75,172} = 16.8\%$. According to Eqn.~\ref{eqn-MPGe-IPC-pc} this implies that the effective powertrain efficiency of PAC-Car II is $\eta_p = 0.168$ (or 16.8\%; note that there is no idling fuel consumption in this case, i.e., $C_{v0} = 0$).

\subsection*{Mileage efficiency of cars}

\begin{figure*}[!t]
\centering
\includegraphics[width=0.7\textwidth]{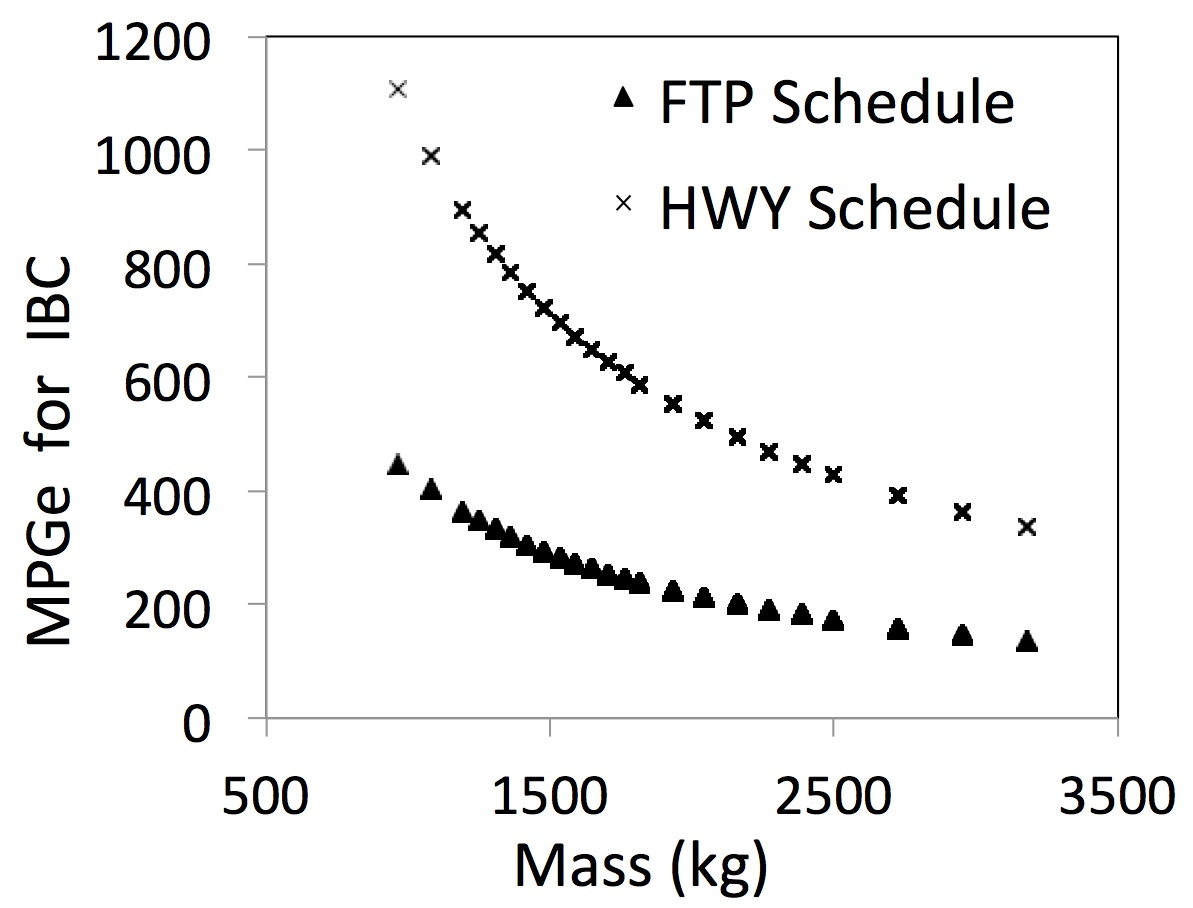}
\caption{Mileage (MPGe) of an ideal brake-loss car for different U.S. EPA driving schedules~\cite{drivesched}.}
\label{fig-MPGeIBCvsMass}
\end{figure*}

\begin{figure*}[!t]
\centering
\includegraphics[width=0.7\textwidth]{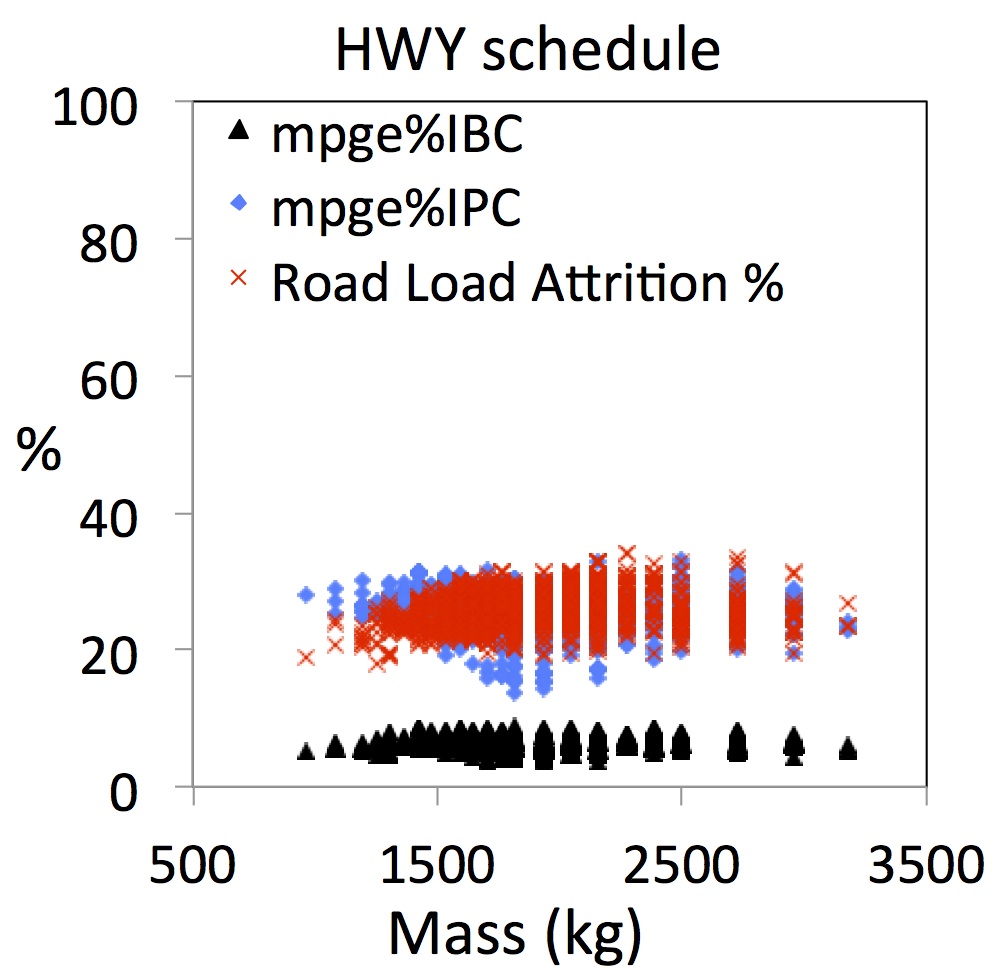}
\includegraphics[width=0.7\textwidth]{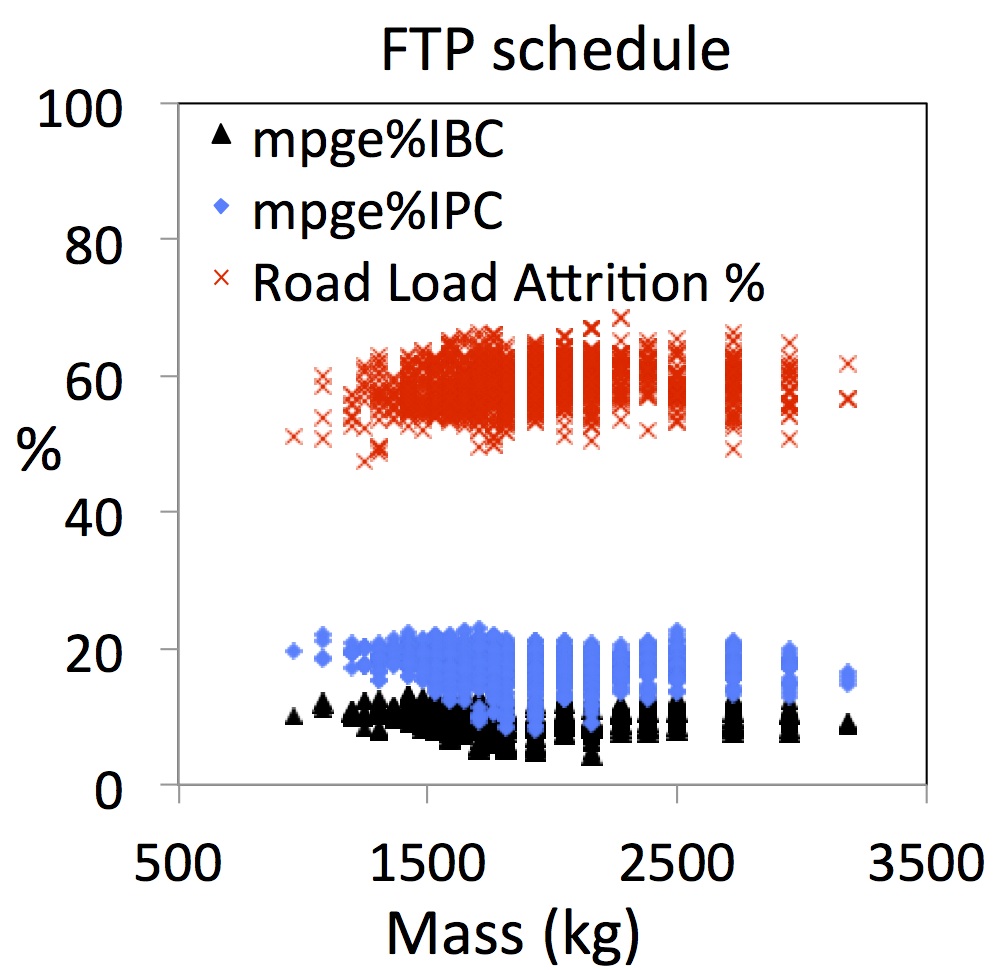}
\caption{Mileage efficiency of 2014 gasoline cars (2014 car mileage data from~\cite{cardata2014}) as a function of car mass and driving schedules.}
\label{fig-MPGEffvsMass}
\end{figure*}

Figure~\ref{fig-MPGeIBCvsMass} shows MPGe of the ideal brake-loss car for different masses and driving schedules. This graph represent the mileage potential (upper limit) of a conventional car. 
What percent of this ideal mileage is actually attained by real-world conventional cars? This question is answered by the mileage efficiency metric $\textup{MPGe}_{\%IBC}$ defined in Eqn.~\ref{eqn-MPGe-IBC-pc} and plotted in Figure~\ref{fig-MPGEffvsMass}.

First it is seen that comparison across mileages of real cars relative to a well defined upper limit has become possible. For example, Figure~\ref{fig-MPGeIBCvsMass} implies that a 2000 kg car giving a mileage of 20 MPGe is almost 10 times less than the ideal mileage for the same mass under urban driving schedule. It is noted that thermodynamic limits due to operating conditions, e.g. Carnot efficiency, or practical efficiency limits of internal combustion engines can lower the practically attainable mileage for a particular technology below the ideal limit. 
However, ideal mileage provides a reference value with which the the mileage of real-world cars can be compared. This informs us how well that machine is functioning relative to the ideal limit.

Second, Figure~\ref{fig-MPGeIBCvsMass} shows ways of increasing the mileage of conventional cars. It shows that reducing mass can increase mileage significantly -- a result that is well known~\cite{Sovr83a, chea10a, Band08a, Chus12a, Knit12a}. Figure~\ref{fig-MPGeIBCvsMass} also shows that the maximum mileage for the highway schedule is more than twice the maximum mileage for the urban schedule. Minor driving changes while beneficial may not significantly improve mileage. 
While car manufacturers could develop light-weight vehicles to increase mileage, there is also a burden on road infrastructure development so that highway-style driving is facilitated to increase vehicle mileages. This is because both light weighting and highway driving can potentially increase mileage by approximately same factor as seen in Figure~\ref{fig-MPGeIBCvsMass}.    

Third, Figure~\ref{fig-MPGEffvsMass} shows that all real world light motor vehicles are, on an average, nearly equally efficient at a particular driving schedule irrespective of the mass of the vehicle. For example, the mileage efficiency is nearly 6.6\% for highway schedule and 11\% for urban schedule for all cars. There is no major change in efficiency with respect to mass. It is evident from this result that a comparison between efficiencies of two cars of different masses has become possible by using the mileage efficiency metric. A comparison between mileages of two cars of different masses gives no insight into which of the two is a more efficiently designed machine.  

Fourth, further insights into the causes of inefficiencies -- internal or external losses -- are possible by using two additional metrics ($\textup{MPGe}_{\%IPC}$ and RLA) defined in this work. These metrics are plotted in Figure~\ref{fig-MPGEffvsMass}. It is generally believed that highway driving is preferred to get high mileage -- this is often interpreted as a more ``efficient" way to drive. While it is true that highway mileage of real cars is higher than that during urban driving, Figure~\ref{fig-MPGEffvsMass} shows that in fact the mileage efficiency is less during the highway schedule (6.6\%) compared to the urban schedule (11\%). As noted in Eqn.~\ref{eqn-MPGe-IBC-IPC-pc}, $\textup{MPGe}_{\%IBC}$ reduces (ideal value is 100\%) due to internal losses quantified by $\textup{MPGe}_{\%IPC}$ and external losses quantified by RLA. Figure~\ref{fig-MPGEffvsMass} shows that $\textup{MPGe}_{\%IPC}$ is nearly 27\% for highway driving and 19\% for urban driving. This means that internal losses are less during highway driving (i.e. the powertrain is functioning more efficiently) compared to urban driving. Yet, the overall mileage efficiency during highway driving is lower because RLA is 24\% for highway driving compared to 57\% for urban driving. Note that a lower value of RLA means greater road loss. This factor of two difference in RLA for highway and urban schedule is because there is substantially higher road load at higher speeds that are typical in highway driving. These results imply that the high mileage potential of highway schedule is not efficiently utilized by current cars due to large road load at high speeds. This emphasizes the importance of tyre and aerodynamic design to tap into the high mileage potential of highway-type schedules. 

It is noted that the EPA states a fuel power-to-wheels efficiency of 14\%-20\% for urban driving and 22\%-30\% for highway driving \cite{energybudget, bagl07a, bagl07b, Band08a}. This is equivalent to the $\textup{MPGe}_{\%IPC}$ metric defined here. The EPA values are consistent with the values of $\textup{MPGe}_{\%IPC}$ in Figure~\ref{fig-MPGEffvsMass}. However, it is clear from the discussion above that this does not translate to mileage efficiency; to do that RLA should be accounted for which is done in this work.


    
Finally, it is noted that regenerative braking and fuels with better energy conversion efficiency can help achieve mileages beyond the ideal limit for conventional cars. 

\subsection*{Mileage efficiency by fuel type}

\begin{figure*}[!t]
\centering
\includegraphics[width=0.7\textwidth]{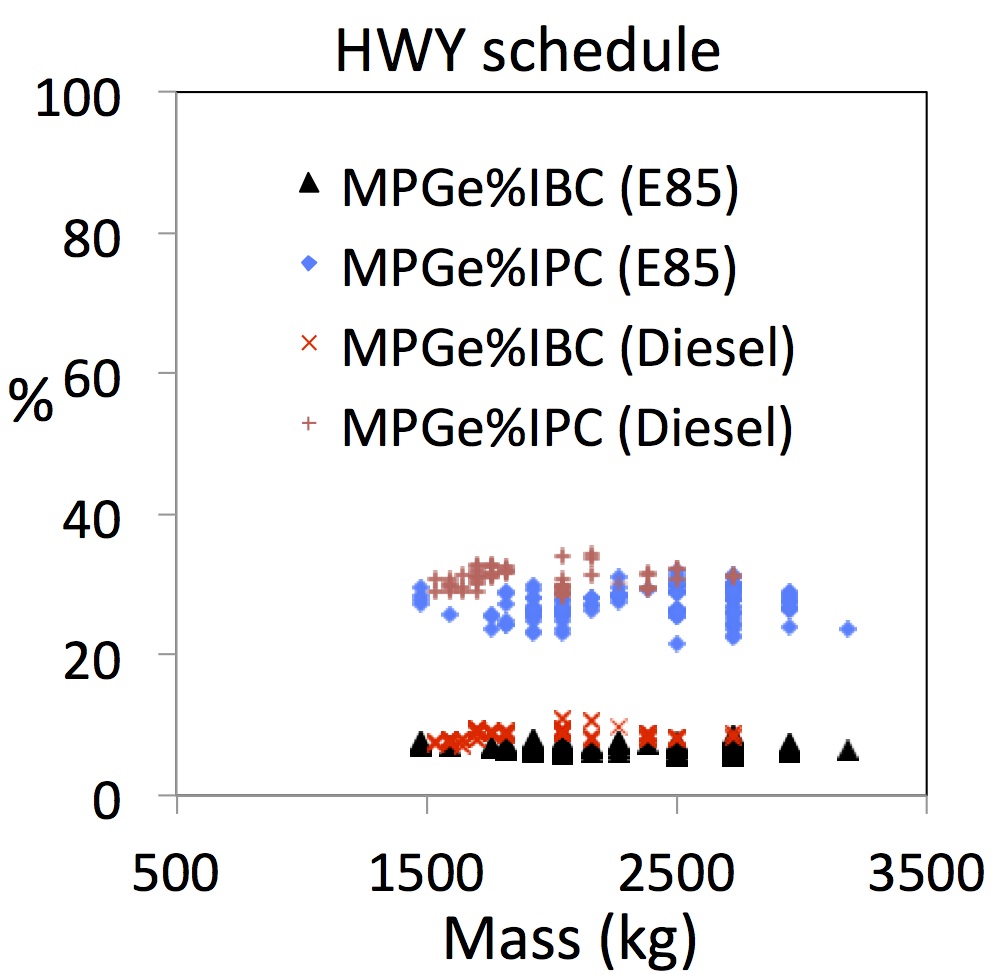}
\includegraphics[width=0.7\textwidth]{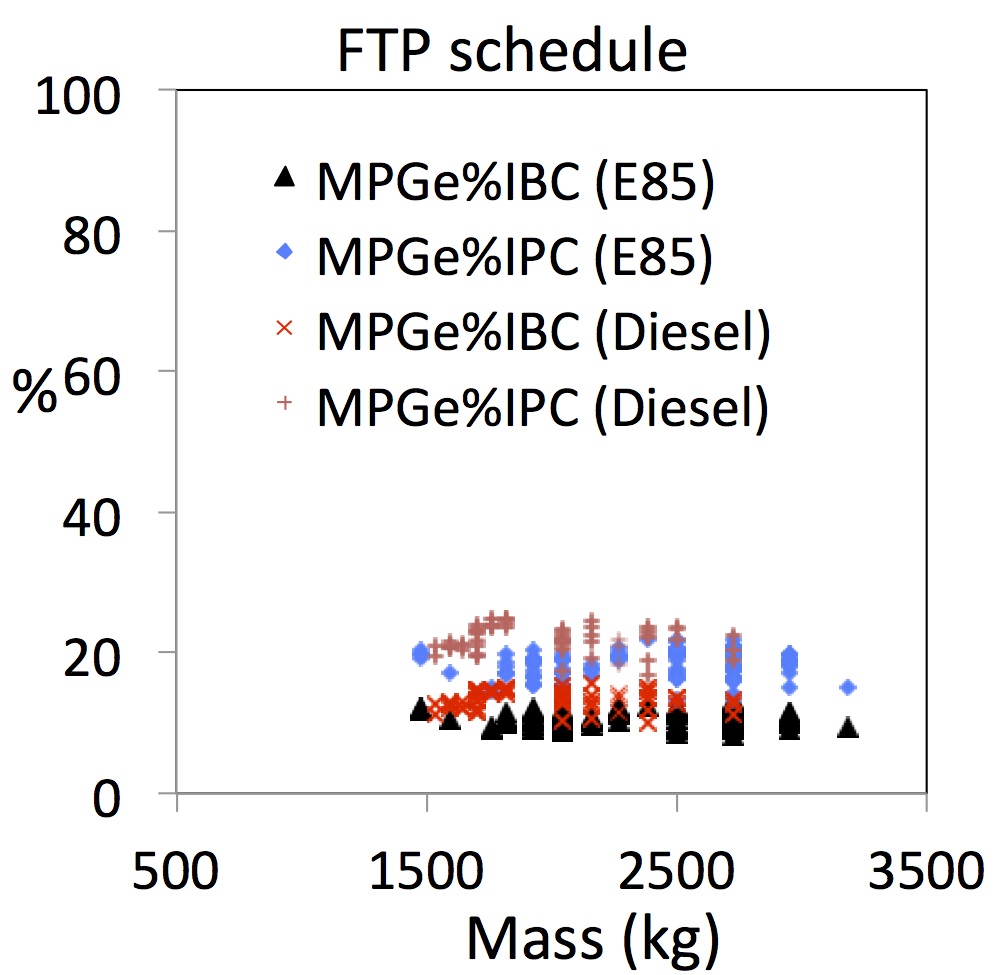}
\caption{Mileage efficiency of 2014 diesel and 85\% ethanol cars (2014 car mileage data from~\cite{cardata2014}) as a function of car mass and driving schedules.}
\label{fig-MPGEffvsMassFuelType}
\end{figure*}

The mileage efficiency metrics defined earlier can be used to check whether cars with a particular fuel type are significantly better than the other. Figure~\ref{fig-MPGEffvsMassFuelType} shows mileage efficiencies for diesel and 85\% ethanol fuel (Figure~\ref{fig-MPGEffvsMass} shows these results for gasoline). It is noted that there is no major  difference in mileage efficiencies between cars using gasoline, diesel, or 85\% ethanol. 

\subsection*{Relative $\textup{CO}_2$ emission by fuel type} 

\begin{figure*}[!t]
\centering
\includegraphics[width=0.7\textwidth]{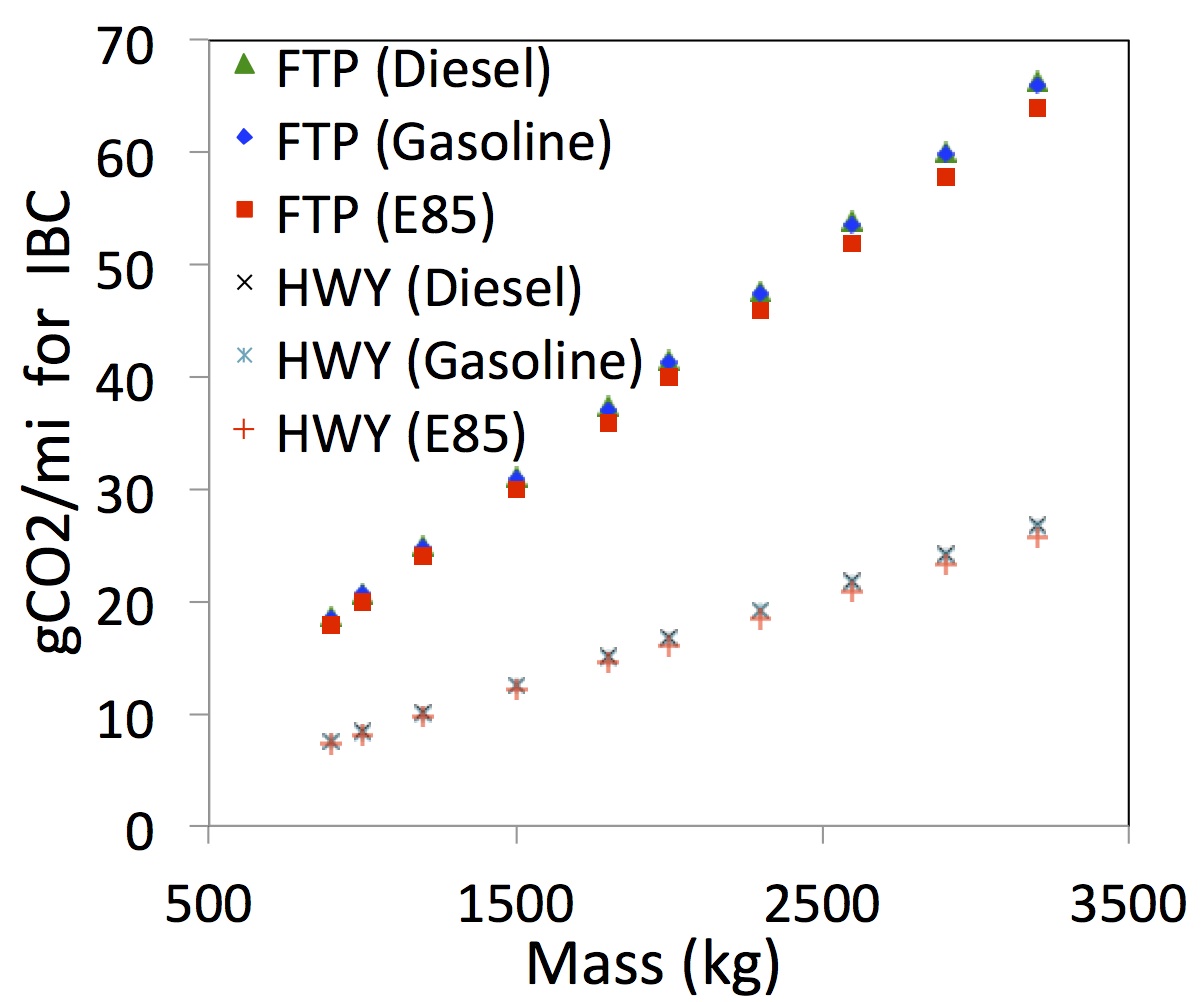}
\caption{$\textup{CO}_2$ emission of an ideal brake-loss car vs. mass for different driving schedules and fuel types.}
\label{fig-gCO2pmiFueltype}
\end{figure*}

\begin{figure*}[!t]
\centering
\includegraphics[width=0.7\textwidth]{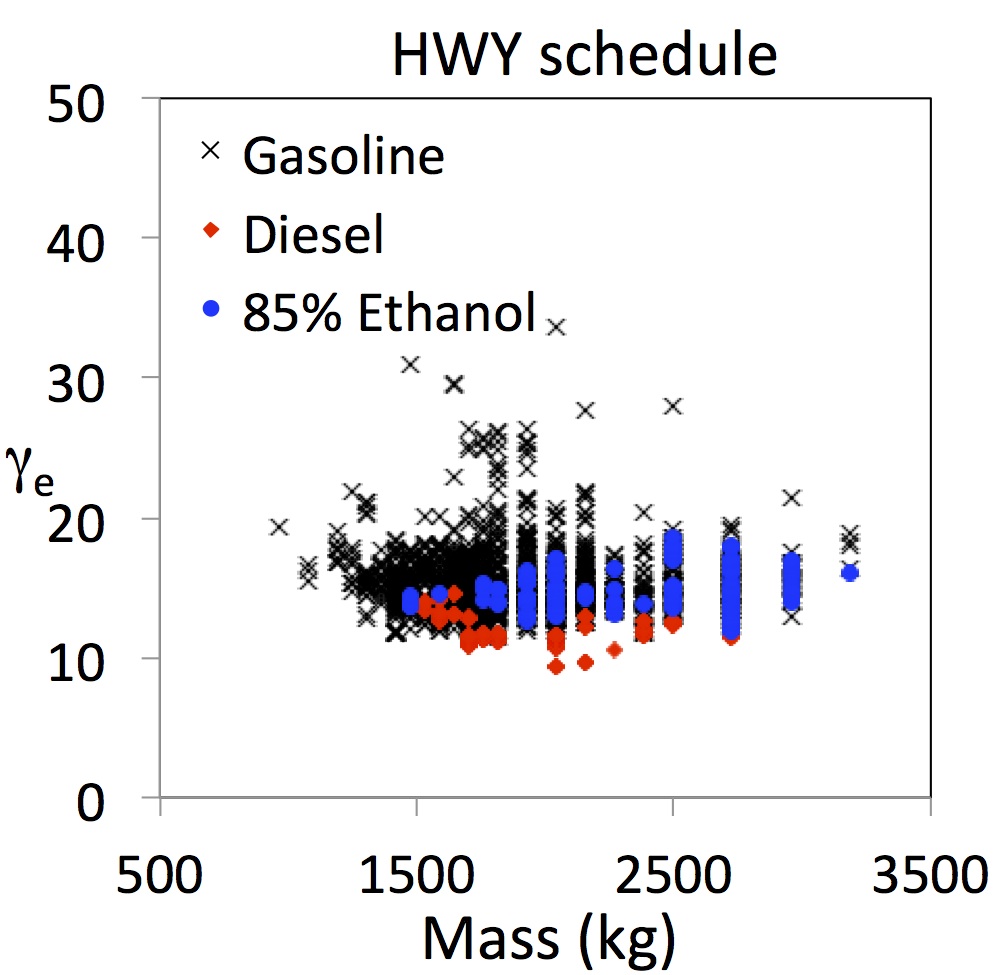}
\includegraphics[width=0.7\textwidth]{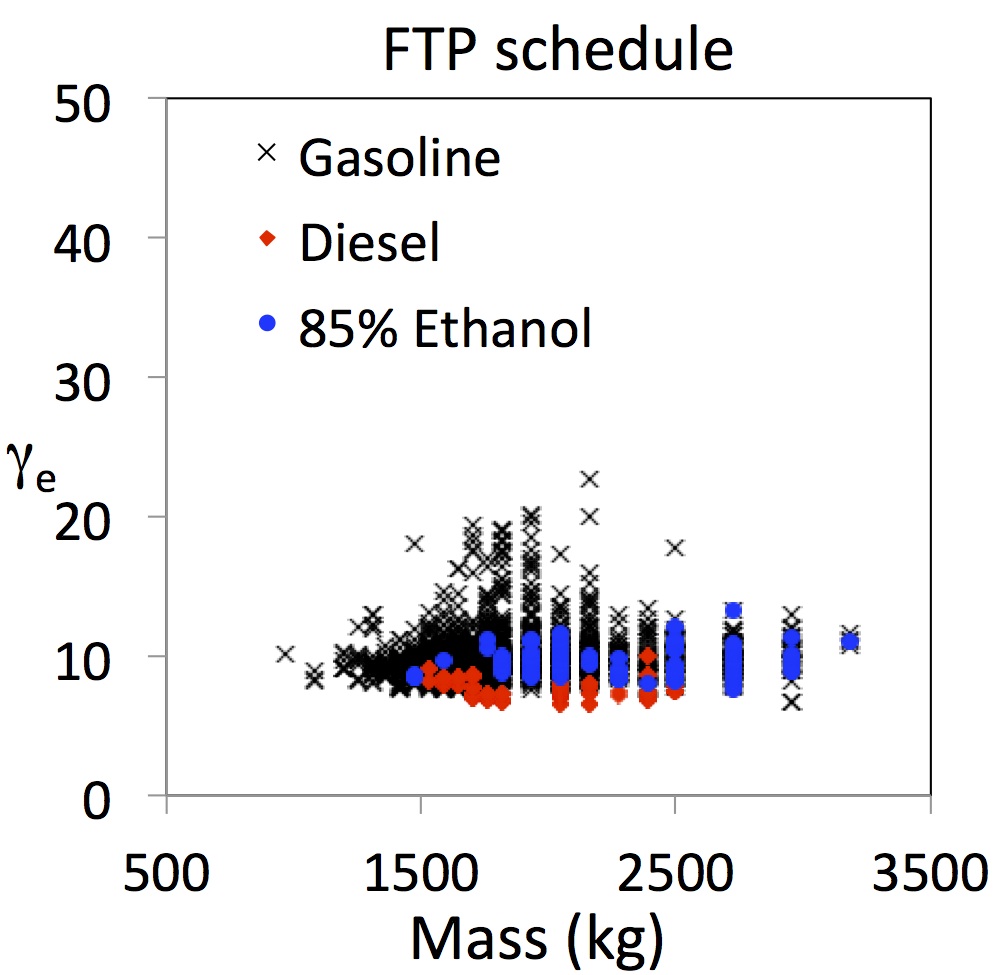}
\caption{Relative $\textup{CO}_2$ emission vs. mass of 2014 cars for different driving schedules and fuel types (2014 car emission data from~\cite{cardata2014}).}
\label{fig-RelEmission}
\end{figure*}

Figure~\ref{fig-gCO2pmiFueltype} shows $\textup{CO}_2$ emission of an ideal brake-loss car (Eqn.~\ref{eqn-ideal-emission}) for gasoline, diesel, and 85\% ethanol. It is seen that 85\% ethanol has the potential for lowest $\textup{CO}_2$ emission, but only marginally, among the three fuel types tested. This is because of the lowest carbon content in 85\% ethanol (i.e. low $\Gamma_{idl}$) among the fuel types considered here. The highway schedule leads to low $\textup{CO}_2$ emission for all fuel types due to greater mileage.

Figure~\ref{fig-RelEmission} (Eqn.~\ref{eqn-relative-emission}) shows the actual to ideal emission ratio of real-world cars for each fuel type. The relative emission from diesel is least and that from 85\% ethanol is intermediate. Gasoline cars have the maximum range and some of the highest values of relative emission $\gamma_e$. Specifically, Figure~\ref{fig-RelEmission} shows that diesel and 85\% ethanol cars emit 6-20 times more $\textup{CO}_2$ per mile compared to ideal values depending on the driving schedule. Gasoline cars emit 8-35 times more $\textup{CO}_2$ per mile compared to ideal values, which is higher than diesel and 85\% ethanol cars. 

Eqn.~\ref{eqn-relative-emission} shows that $\gamma_e$ depends on $C_E$ (which is inversely proportional to mileage efficiency) and $\eta_r$. It is found that $\eta_r$ values are close to one. This indicates that even if the combustion is not perfect in real-world cars, the catalytic convertors are very efficient. Thus, the relative emission is almost entirely dependent on the value of $C_E$.

\subsection*{Historical trend of mileage efficiency by manufacturers} 

\begin{figure*}[!t]
\centering
\includegraphics[width=0.7\textwidth]{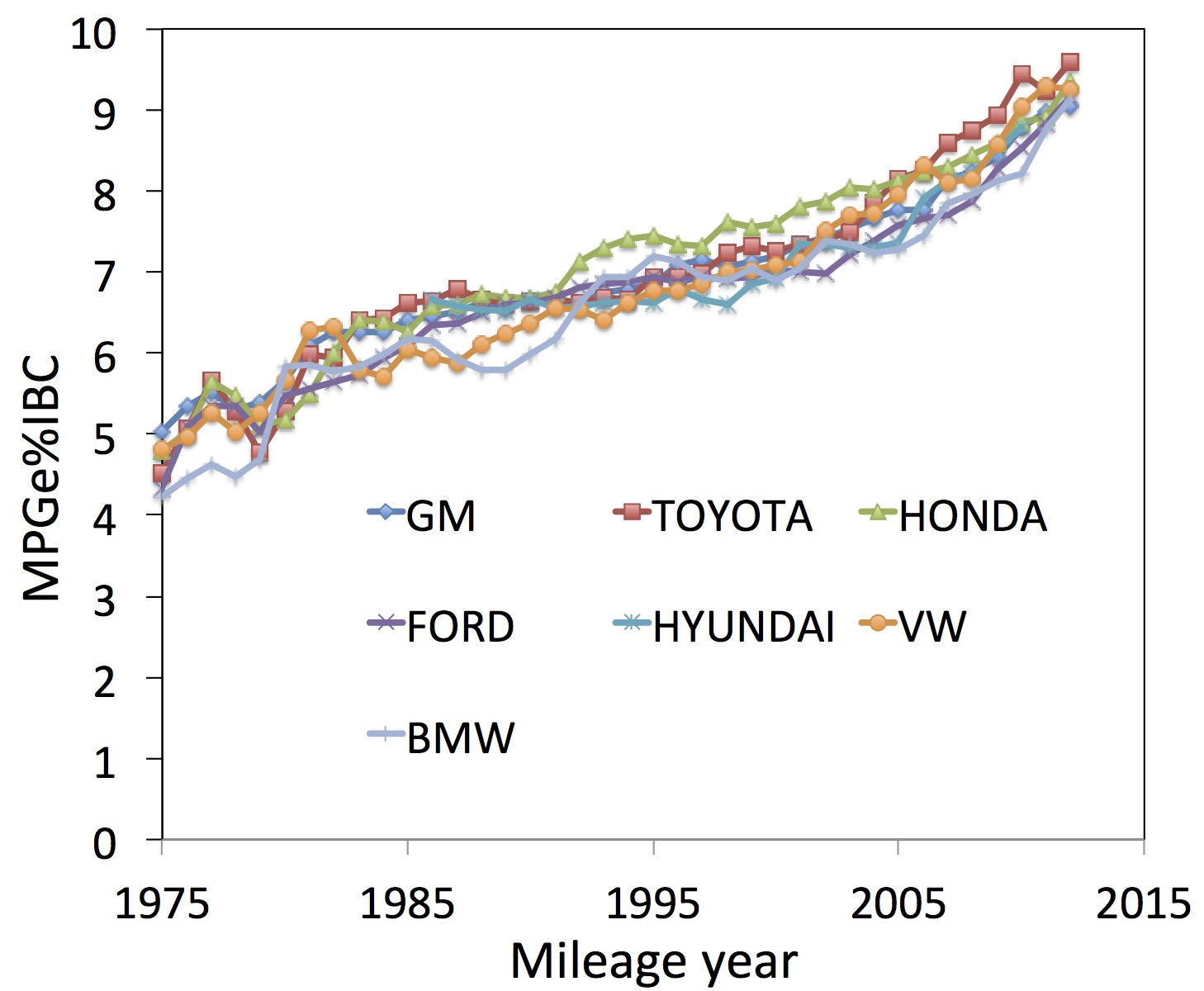}
\caption{A historical trend of mileage efficiency of cars for different manufacturers (historical mileage data from~\cite{manutrend}).}
\label{fig-MPGEffManufacturer}
\end{figure*}

Figure~\ref{fig-MPGEffManufacturer} shows historical trend for mileage efficiency for different car manufacturers. The average mileage efficiency has steadily risen from around 4.5\% to more than 9\% at present. It is also noted that all manufacturers' performance is clustered without significant difference. 

\subsection*{Setting fuel economy and emission standards}

The mileage efficiency and relative emission metrics can be a useful tool to set uniform emissions standards around the world independent of the driving pattern in a particular country. Setting limits based on these metrics will encourage car manufacturers to design cars that efficiently tap into the ideal mileage potential for a particular fuel type.




\section*{Methods}

All data were obtained from public domain databases of U.S. EPA. Urban and highway driving schedule data (velocity vs. time) were obtained from~\cite{drivesched}. The actual mileage, emission, and road load data were obtained from~\cite{cardata2014}. Historical fuel economy data for manufacturers were obtained from~\cite{manutrend}. 

\bibliography{./references}

\bibliographystyle{Science}

\end{document}